\documentclass[aps,twocolumn,pre,amsmath,amssymb]{revtex4}
\usepackage{graphicx}
\usepackage{epstopdf}
\usepackage{pstricks}
\newcommand{\be}{\begin{equation}}
\newcommand{\ee}{\end{equation}}
\newcommand{\ba}{\begin{array}{l}}
\newcommand{\ea}{\end{array}}
\newcommand{\re}[1]{(\ref{#1})}
\newcommand{\ci}[1]{\cite{#1}}
\newcommand{\banonum}{\begin{eqnarray*}}
\newcommand{\eanonum}{\end{eqnarray*}}
\newcommand{\baa}{\begin{eqnarray}}
\newcommand{\eaa}{\end{eqnarray}}
\newcommand{\bfr}{\begin{flushright}}
\newcommand{\efr}{\end{flushright}}
\newcommand{\bfl}{\begin{flushleft}}
\newcommand{\efl}{\end{flushleft}}
\newcommand{\lab}[1]{\label{#1}}
\begin{document}
\title{Bogoliubov de Gennes equation on metric graphs}

\author{K.K. Sabirov$^a$, D. Jumanazarov$^b$, J.Yusupov$^c$, D.U. Matrasulov$^c$}

\affiliation{$^a$Tashkent University of  Information Technology, Amir Temur Avenue 108, Tashkent 100200, Uzbekistan\\
$^b$Institut d'Alembert, ENS Paris-Saclay University, 61-avenue du President Wilson 94230, Cachan, France\\
$^c$ Turin Polytechnic University in Tashkent, 17 Niyazov str., 100095,
Tashkent, Uzbekistan}

\begin{abstract}
We consider Bogoliubov de Gennes equation on metric graphs. The
vertex boundary conditions  providing self-adjoint realization of
the Bogoliubov de Gennes operator on a metric star graph  are
derived. Secular equation providing quantization of the energy and
the vertex transmission matrix are also obtained. Application of
the model for Majorana wire networks is discussed.
\end{abstract}
\maketitle

\section{Introduction}\lab{intro}

Quantum graphs, which are the one- or quasi-one dimensional
branched quantum wires, have attracted much attention in different
contexts of contemporary physics (see, e.g.,
Refs.\ci{Uzy1}-\ci{Exner15}). Particle dynamics in such systems
are described in terms of quantum mechanical wave equations on
metric graphs by imposing the boundary conditions at the branching
points (vertices) and bond ends. The metric graphs are the set of
bonds which are assigned length and connected to each other at the
vertices. The connection rule is called topology of a graph and
given in terms of the adjacency matrix \ci{Uzy1,Uzy2}.

Quantum graphs were first introduced by Exner and Seba to describe
free quantum motion on branched structures\ci{Exner1}. Later
Kostrykin and Schrader derived the general boundary conditions
providing self-adjointness of the Schr\"odinger operator on graphs
\ci{Kost}. Bolte and Harrison extended such boundary conditions
for the Dirac operator on metric graphs \ci{Bolte}. Hul \emph{et
al} considered experimental realization of quantum graphs in
optical microwave networks \ci{Hul}. An important topic related to
quantum graphs was studied in the context of quantum chaos theory
and spectral statistics \ci{Uzy1,Bolte,Uzy2,Uzy3,Harrison}.
Spectral properties and band structure of periodic quantum graphs
also attracted much interest \ci{Grisha1,Kotlyarov}. Different
aspects of the Schr\"odinger operators on graphs have been studied
in the Refs.\ci{Exner15,Grisha,Mugnolo,Rabinovich,Bolte1}.

Despite the growing interest to quantum graphs which are described
by linear wave equation on metric graphs, within such approach one
is restricted by modeling linear wave dynamics only. For modeling
of  nonlinear waves and soliton dynamics in branched structures
one should consider nonlinear wave equations on metric graphs.
During the past decade the studies of particle and wave dynamics
in branched structures have been extended to nonlinear evolution
equations by considering nonlinear Schr\"odinger and sine-Gordon
equations on metric graphs \ci{Zarif}-\ci{Our1}. For such
equations, one should derive the vertex boundary conditions from
fundamental conservation laws such as energy, momentum, charge and
mass conservation \ci{Zarif,Our1}. Due to the numerous
applications of metric graphs based approach for wave dynamics in
branched systems and networks, one can expect further extension of
the studies to the case of other evolution equations.

In this paper we consider quantum graphs described by Bogoliubov de Gennes
(BdG) equation. The latter can be used for modeling of quasiparticle dynamics
in superconductors \ci{Gennes} and Majorana fermions  in superconducting
quantum wires \ci{Jackiw}. Here we derive the vertex boundary conditions which
keep the BdG operator on metric graphs as self-adjoint. Explicit solutions for
such boundary conditions are obtained. Also, we derive vertex transmission
matrix describing of waves through the graph branching point.

The paper is organized as follows. In the next section we give formulation of
the problem and derive vertex boundary conditions providing self-adjointness of
the problem. Section \ref{vt} presents derivation of the transmission matrix
describing wave transmission through the graph branching point. In section
\ref{mw} we discuss zero-mode solutions of BdG equation on metric graph and
possible application to Majorana wire networks. Finally, section \ref{conc}
presents some concluding remarks.

\section{Vertex boundary conditions and explicit solutions}\lab{vbc}

First order BdG equation which is often used in condensed matter physics can be
written as
\be
 H_\text{BdG}\Psi = E\Psi,
\label{bdg1}
 \ee
where $\Psi = \left(\Psi_{1}, \Psi_2, \Psi_3, \Psi_4\right)^T$ and
\begin{align}
H_\text{BdG} = \left(\begin{array}{cccc}
0&-i\frac{\partial}{\partial x}&\Delta_0&0\\
-i\frac{\partial}{\partial x}&0&0&\Delta_0\\
\Delta_0&0&0&i\frac{\partial}{\partial x}\\
0&\Delta_0&i\frac{\partial}{\partial x}&0\end{array}\right) \label{eq2}
\end{align}

General solution of Eq.(\ref{bdg1}) can be written as
\begin{eqnarray}
\Psi_1(x,E)&=&C_{11}e^{i\kappa x}+C_{12}e^{-i\kappa x},\nonumber\\
\Psi_2(x,E)&=&C_{21}e^{i\kappa x}+C_{22}e^{-i\kappa x},\nonumber\\
\Psi_3(x,E)&=&\frac{E}{\Delta_0}C_{11}e^{i\kappa
x}+\frac{E}{\Delta_0}C_{12}e^{-i\kappa x}\nonumber\\
&-&\frac{\kappa}{\Delta_0}C_{21}e^{i\kappa x}+\frac{\kappa}{\Delta_0}C_{22}e^{-i\kappa x},\nonumber\\
\Psi_4(x,E)&=&-\frac{\kappa}{\Delta_0}C_{11}e^{i\kappa
x}+\frac{\kappa}{\Delta_0}C_{12}e^{-i\kappa x}\nonumber\\
&+&\frac{E}{\Delta_0}C_{21}e^{i\kappa x}+\frac{E}{\Delta_0}C_{22}e^{-i\kappa
x},\label{eq3}
\end{eqnarray}
where $\kappa=\sqrt{E^2-\Delta_0^2}$ and $C_{11}, C_{12}, C_{21}, C_{22}$ are
constants which can be found, e.g. from the normalization and boundary
conditions. Current for such system is determined as \be J =
\Psi^*\left(\begin{array}{cc}\sigma_x&0\\0&\sigma_x\end{array}\right)\Psi.\lab{current}\ee

\begin{figure}[ht!]
\includegraphics[width=60mm]{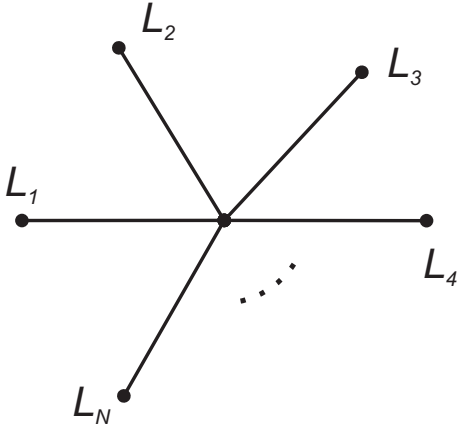}
\caption{Sketch of a metric star graph. $L_j$ is the length of the $j$th bond
with $j=1,2,\ldots, N$.} \label{pic1}
\end{figure}

Here we address the problem of BdG equation on metric graphs by considering a
metric star graph with $N$ bonds,  having the finite lengths,
$L_j,\,j=1,2,\ldots,B$ (see, Fig.1). On each bond we we have the following BdG
equation: \be H_\text{BdG}\Psi^{(j)} = E\Psi^{(j)}, \label{bdg2} \ee where
$j=1,2,\ldots,N$ is the bond number, $\Psi^{(j)}$ and $H_\text{BdG}$ are the
wave function and BdG operator (respectively), which are given by $\Psi^{(j)} =
\left(\Psi_1^{(j)}, \Psi_2^{(j)}, \Psi_3^{(j)}, \Psi_4^{(j)}\right)^T$ and
\begin{align}
H_\text{BdG} = \left(\begin{array}{cccc}0&-i\frac{\partial}{\partial
x}&\Delta_0&0\\-i\frac{\partial}{\partial
x}&0&0&\Delta_0\\\Delta_0&0&0&i\frac{\partial}{\partial
x}\\0&\Delta_0&i\frac{\partial}{\partial x}&0\end{array}\right) \label{eq2}
\end{align}

Eq.\re{bdg2} describes quasiparticle dynamics in branched superconductors
\ci{Kos,Serret,Mancini} and (for $E=0$) Majorana wire networks
\ci{Clarke,Alicea1,Halperin,Pedrocci}. The latter has attracted much attention
in the context of topological quantum computation recently
\ci{Pedrocci,KB,Hell}. For simplicity, in the following we assume that
$\Delta_0=\text{const}$. To solve Eq.\re{bdg2} on metric graph, we need to
impose the boundary conditions at the graph vertex. Such boundary conditions
should not break self-adjointness of the BdG operator. Here we use the
prescription proposed first in \ci{Kost} for the Schrodinger equation on metric
graph and developed later for Dirac equation in \ci{Bolte}.  We construct the
scalar product of two functions on the graph, $\phi$ and $\psi$ which given by
\begin{align}
\langle\phi,\psi\rangle=\sum_{j=1}^N\sum_{k=1}^4\int_{0}^{L_k}\phi_k^{(j)}(x)\bar\psi_k^{(j)}(x)dx.
\end{align}
Also, for a given differential operator, $D$ on a graph we define so-called
skew-Hermitian form which is given as \ci{Kost,Bolte}
\begin{widetext}
\begin{align}
\Omega(\psi,\phi)=&\langle
D\psi,\phi\rangle-\langle\psi,D\phi\rangle\nonumber\\
=&i\sum_{j=1}^N\left(-\psi_{2}^{(j)}(L_j)\bar\phi_{1}^{(j)}(L_j)-\psi_{1}^{(j)}(L_j)\bar\phi_{2}^{(j)}(L_j)+\psi_{4}^{(j)}(L_j)\bar\phi_{3}^{(j)}(L_j)+\psi_{3}^{(j)}(L_j)\bar\phi_{4}^{(j)}(L_j)\right.\nonumber\\
&\left.+\psi_{2}^{(j)}(0)\bar\phi_{1}^{(j)}(0)+\psi_{1}^{(j)}(0)\bar\phi_{2}^{(j)}(0)-\psi_{4}^{(j)}(0)\bar\phi_{3}^{(j)}(0)-\psi_{3}^{(j)}(0)\bar\phi_{4}^{(j)}(0)\right)
\end{align}
Using the notations
\begin{align}
&\psi_1:=\left(\psi_1^{(1)}(0);...;\psi_1^{(N)}(0);\psi_3^{(1)}(0);...;\psi_3^{(N)}(0);\psi_1^{(1)}(L_1);...;\psi_1^{(N)}(L_N);\psi_3^{(1)}(L_1);...;\psi_3^{(N)}(L_N)\right)^T,\nonumber\\
&\psi_2:=\left(\psi_2^{(1)}(0);...;\psi_2^{(N)}(0);-\psi_4^{(1)}(0);...;-\psi_4^{(N)}(0);-\psi_2^{(1)}(L_1);...;-\psi_2^{(N)}(L_N);\psi_4^{(1)}(L_1);...;\psi_4^{(N)}(L_N)\right)^T,\label{eq2}
\end{align}
\end{widetext}
for  $\Omega$ we have
\begin{align}
\Omega(\psi,\phi)=i\left(\begin{array}{cc}\phi_1^\dagger&\phi_2^\dagger\end{array}\right)\left(\begin{array}{cc}0&I_{4N}\\
I_{4N}&0\end{array}\right)\left(\begin{array}{cccc}\psi_1\\
\psi_2\end{array}\right).\label{eq3}
\end{align}
Then the vertex boundary conditions can be written in a linear subspace of
${\bf C}^{8N}$ as \ci{Kost,Bolte}
\begin{align}
{\bf A}\psi_1+{\bf B}\psi_2=0, \label{eq4}
\end{align} with complex $4N\times4N$ matrices ${\bf A}$ and ${\bf B}$.
Using the relations \ci{Bolte}
\begin{align}
\psi_1=-{\bf A}^{-1}{\bf B}\psi_2,\,\,\,\,\phi_1^\dagger=-\phi_2^\dagger{\bf
B}^\dagger({\bf A}^{-1})^\dagger\nonumber
\end{align}
we get
\begin{align}
\Omega(\psi,\phi)&=iu\left(\begin{array}{cc}\phi_2^\dagger&\phi_2^\dagger\end{array}\right)\nonumber\\
&\cdot\left(\begin{array}{cc}0&-{\bf B}^\dagger({\bf A}^{-1})^\dagger\\-{\bf
A}^{-1}{\bf
B}&0\end{array}\right)\left(\begin{array}{c}\psi_2\\\psi_2\end{array}\right)\label{eq5}
\end{align}
Then we have from $\Omega(\phi,\psi)=0$
\begin{align}
\text{rank}({\bf A})=\text{rank}({\bf B})=4N,\label{rank}
\end{align}
\begin{align}
{\bf A}{\bf B}^\dagger=-{\bf B}{\bf A}^\dagger.\label{condAB}
\end{align}
Solution of Eq.\re{bdg2} for the positive energy ($H_\text{BdG}\Psi^{(j)} =
E\Psi^{(j)}$) can be written as
\begin{align}
\Psi^{(j)}(x)=&\mu_\alpha^{(j)}\left(\begin{array}{c}1\\0\\\frac{E}{\Delta_0}\\-\frac{\kappa}{\Delta_0}\end{array}\right)e^{i\kappa x}
+\mu_\beta^{(j)}\left(\begin{array}{c}0\\1\\-\frac{\kappa}{\Delta_0}\\\frac{E}{\Delta_0}\end{array}\right)e^{i\kappa x}\nonumber\\
+&\hat{\mu}_\alpha^{(j)}\left(\begin{array}{c}1\\0\\\frac{E}{\Delta_0}\\\frac{\kappa}{\Delta_0}\end{array}\right)e^{-i\kappa
x}
+\hat{\mu}_\beta^{(j)}\left(\begin{array}{c}0\\1\\\frac{\kappa}{\Delta_0}\\\frac{E}{\Delta_0}\end{array}\right)e^{-i\kappa
x}.\label{eq6}
\end{align}
For the negative energy ($H_\text{BdG}\Psi^{(j)} = -E\Psi^{(j)}$) we have
\begin{align}
\Psi^{(j)}(x)=&\mu_\alpha^{(j)}\left(\begin{array}{cccc}-\frac{E}{\Delta_0}\\\frac{\kappa}{\Delta_0}\\1\\0\end{array}\right)e^{i\kappa x}+\mu_\beta^{(j)}\left(\begin{array}{cccc}\frac{\kappa}{\Delta_0}\\-\frac{E}{\Delta_0}\\0\\1\end{array}\right)e^{i\kappa x}\nonumber\\
+&\hat{\mu}_\alpha^{(j)}\left(\begin{array}{cccc}-\frac{E}{\Delta_0}\\-\frac{\kappa}{\Delta_0}\\1\\0\end{array}\right)e^{-i\kappa
x}+\hat{\mu}_\beta^{(j)}\left(\begin{array}{cccc}-\frac{\kappa}{\Delta_0}\\-\frac{E}{\Delta_0}\\0\\1\end{array}\right)e^{-i\kappa
x}.\label{eq7}
\end{align}
For these solutions the vertex boundary conditions given by Eq.(\ref{eq4}) can
be written as (for $E > 0 $)
\begin{equation}
\left({\bf A}\Theta_1+{\bf B}\Theta_2\right)
\left(\begin{array}{c}\mu_\alpha\\\mu_\beta\\\hat{\mu}_\alpha\\\hat{\mu}_\beta\end{array}\right)=0\label{eq8}
\end{equation}
where
$$
\Theta_1=\left(\begin{array}{cccc}I_N&0&I_N&0\\\frac{E}{\Delta_0}I_N&-\frac{\kappa}{\Delta_0}I_N&\frac{E}{\Delta_0}I_N&\frac{\kappa}{\Delta_0}I_N\\e^{i\kappa
L}&0&e^{-i\kappa L}&0\\\frac{E}{\Delta_0}e^{i\kappa
L}&-\frac{\kappa}{\Delta_0}e^{i\kappa L}&\frac{E}{\Delta_0}e^{-i\kappa
L}&\frac{\kappa}{\Delta_0}e^{-i\kappa L}\end{array}\right),
$$
$$
\Theta_2=\left(\begin{array}{cccc}0&I_N&0&I_N\\\frac{\kappa}{\Delta_0}I_N&-\frac{E}{\Delta_0}I_N&-\frac{\kappa}{\Delta_0}I_N&-\frac{E}{\Delta_0}I_N\\0&-e^{i\kappa
L}&0&e^{-i\kappa L}\\-\frac{\kappa}{\Delta_0}e^{i\kappa
L}&\frac{E}{\Delta_0}e^{i\kappa L}&\frac{\kappa}{\Delta_0}e^{-i\kappa
L}&\frac{E}{\Delta_0}e^{-i\kappa L}\end{array}\right),
$$
and (for $E < 0$)
\begin{equation}
\left({\bf A}\Theta_3+{\bf
B}\Theta_4\right)\left(\begin{array}{c}\mu_\alpha\\\mu_\beta\\\hat{\mu}_\alpha\\\hat{\mu}_\beta\end{array}\right)=0\label{eq9}
\end{equation}
where
$$
\Theta_3=\left(\begin{array}{cccc}-\frac{E}{\Delta_0}I_N&\frac{\kappa}{\Delta_0}I_N&-\frac{E}{\Delta_0}I_N&-\frac{\kappa}{\Delta_0}I_N\\I_N&0&I_N&0\\-\frac{E}{\Delta_0}e^{i\kappa
L}&\frac{\kappa}{\Delta_0}e^{i\kappa L}&-\frac{E}{\Delta_0}e^{-i\kappa
L}&-\frac{\kappa}{\Delta_0}e^{-i\kappa L}\\e^{i\kappa L}&0&e^{-i\kappa
L}&0\end{array}\right),
$$
$$
\Theta_4=\left(\begin{array}{cccc}\frac{\kappa}{\Delta_0}I_N&-\frac{E}{\Delta_0}I_N&-\frac{\kappa}{\Delta_0}I_N&-\frac{E}{\Delta_0}I_N\\0&-I_N&0&-I_N\\-\frac{\kappa}{\Delta_0}e^{i\kappa
L}&\frac{E}{\Delta_0}e^{i\kappa L}&\frac{\kappa}{\Delta_0}e^{-i\kappa
L}&\frac{E}{\Delta_0}e^{-i\kappa L}\\0&e^{i\kappa L}&0&e^{-i\kappa
L}\end{array}\right).
$$

Eqs. \re{eq8} and \re{eq9} leads to quantization conditions for finding the
eigenvalues from the following secular equations:
\begin{equation}
\det\left({\bf A}\Theta_1+{\bf B}\Theta_2\right)=0\label{eq10}
\end{equation}
for the positive energy and
\begin{equation}
\det\left({\bf A}\Theta_3+{\bf B}\Theta_4\right)=0\label{eq11}
\end{equation}
for the negative energy. Here $e^{i\kappa L}=\text{diag}\{e^{i\kappa
L_1},\ldots,e^{i\kappa L_N}\}$,
$\mu_\alpha=\{\mu_\alpha^{(1)},\ldots,\mu_\alpha^{(N)}\}$,
$\mu_\beta=\{\mu_\beta^{(1)},\ldots,\mu_\beta^{(N)}\}$,
$\hat{\mu}_\alpha=\{\hat{\mu}_\alpha^{(1)},\ldots,\hat{\mu}_\alpha^{(N)}\}$,
$\hat{\mu}_\beta=\{\hat{\mu}_\beta^{(1)},\ldots,\hat{\mu}_\beta^{(N)}\}$, $I_N$
is the identity matrix with the $N$th order.

Eqs. \re{eq6}, \re{eq7}, \re{eq10} and \re{eq11} present complete
set of the eigenfunctions and eigenvalues of Eq.\re{bdg2} for the
vertex boundary conditions \re{eq8} and \re{eq9}. It is clear that
these solutions provide current conservation in the form of
kirchhoff rules at the vertex. This can be directly checked from
the definition of the current given by Eq.\re{current}.

\section{Vertex transmission}\lab{vt}

Here we treat the problem of wave transmission through the graph branching
point. Defining the vectors for outgoing and incoming waves at the vertex
\begin{equation}
\overrightarrow{\mu}=\left(\begin{array}{c}\mu_\alpha\\
\mu_\beta\\e^{-i\kappa L}\hat{\mu}_\alpha\\e^{-i\kappa
L}\hat{\mu}_\beta\end{array}\right),\,\overleftarrow{\mu}=\left(\begin{array}{c}\hat{\mu}_\alpha\\
\hat{\mu}_\beta\\e^{i\kappa L}\mu_\alpha\\e^{i\kappa
L}\mu_\beta\end{array}\right),\nonumber
\end{equation}
from the vertex boundary conditions we have
\begin{equation}
\overrightarrow{\mu}=-\left({\bf AA'}+{\bf
BB'}\right)^{(-1)}\left({\bf AA''}+{\bf
BB''}\right)\overleftarrow{\mu},\label{eq12}
\end{equation}
where
$${\bf
A}'=\left(\begin{array}{cccc}I_N&0&0&0\\\frac{E}{\Delta_0}I_N&-\frac{\kappa}{\Delta_0}I_N&0&0\\0&0&I_N&0\\0&0&\frac{E}{\Delta_0}I_N&\frac{\kappa}{\Delta_0}I_N\end{array}\right),
$$
$${\bf
A}''=\left(\begin{array}{cccc}I_N&0&0&0\\\frac{E}{\Delta_0}I_N&\frac{\kappa}{\Delta_0}I_N&0&0\\0&0&I_N&0\\0&0&\frac{E}{\Delta_0}I_N&-\frac{\kappa}{\Delta_0}I_N\end{array}\right),
$$
$${\bf
B}'=\left(\begin{array}{cccc}0&I_N&0&0\\\frac{\kappa}{\Delta_0}I_N&-\frac{E}{\Delta_0}I_N&0&0\\0&0&0&-I_N\\0&0&\frac{\kappa}{\Delta_0}I_N&\frac{E}{\Delta_0}I_N\end{array}\right),
$$
$${\bf
B}''=\left(\begin{array}{cccc}0&I_N&0&0\\-\frac{\kappa}{\Delta_0}I_N&-\frac{E}{\Delta_0}I_N&0&0\\0&0&0&-I_N\\0&0&-\frac{\kappa}{\Delta_0}I_N&\frac{E}{\Delta_0}I_N\end{array}\right)
$$
and $\text{det}\left({\bf AA'}+{\bf BB'}\right)\not=0$. Then the vertex
transition matrix can be written as \ci{Bolte}
\begin{equation}
\bf{T}=-\left({\bf AA'}+{\bf BB'}\right)^{(-1)}\left({\bf
AA''}+{\bf BB''}\right).\label{eq13}
\end{equation}

Following the Refs. \ci{Uzy1,Bolte}, one can construct the bond scattering
matrix in terms of the transmission matrix as
\begin{equation}
S_{(ij)(l0)}=\delta_{0i}T_{(ij)(l0)}e^{i\kappa L_{(l0)}},\label{scatmat1}
\end{equation}
where $(mn)$ is the bond connected vertices $m$ and $n$.

\begin{figure}[ht!]
\includegraphics[width=80mm]{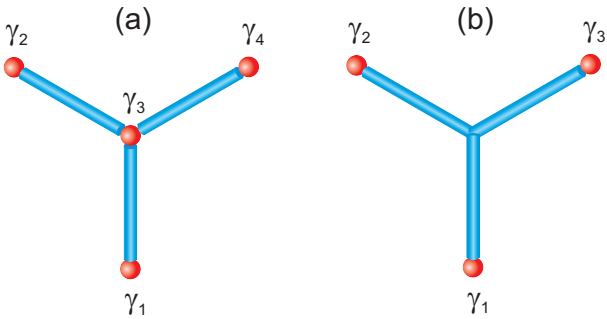}
\caption{(Color online). Sketch of a branched (Y-junction)
Majorana wire. Each $\gamma$ denotes Majorana fermion located at
the end of a branch. } \label{pic1}
\end{figure}

\section{Majorana wire network}\lab{mw}

The above treatment concerns non-zero energy ($E\neq 0$) solutions of BdG
equation on metric graph. An important case having application in topological
states of condensed matter is described by zero-energy solutions of BdG.  Such
solutions describe bound states of  the Majorana fermions on a quantum wire
(so-called Majorana wires) \ci{Jackiw,Alicea1,LF} which are localized at the
ends of the wire. Majorana particles in quantum wires are considered as fixed
(immobile), i.e. they do not carry any current \ci{Jackiw,Alicea1}. However,
one can achieve current carrying regime by constructing Y(T)-junctions of
Majorana wires, or more complicated branching topologies
\ci{Alicea1}-\ci{Hell}. Modeling of such systems can be done in terms of BdG
equation on metric graphs. Thus the problem we want to address is the BdG
equation on metric star graph given by \be H_\text{BdG}\Psi^{(j)} =0,
\label{bdg3} \ee where the spinor $\Psi^{(j)}$ has (for Majorana fermions) the
Nambu structure which is given as \ci{Jackiw,Alicea1}
$$
\Psi^{(j)}(x)=\left(\Psi^{(j)}_{\uparrow},\Psi^{(j)}_{\downarrow},\Psi^{(j)*}_{\downarrow},-\Psi^{(j)*}_{\uparrow}\right)^T.
$$

Eq.\re{bdg3} describes Majorana  wire networks. Such networks have attracted
much attention during last few years ( see, e.g., Refs.\ci{Clarke}-\ci{Hell}).
Typical examples of Majorana wire networks are resented in Fig.2. For T- or
Y-junction of quantum wires two types of branching is possible: First type has
Majorana fermion at the vertex (Fig. 2a), while for in second type vertex does
not contain Majorana fermion (Fig.2b). Using different pairing rules and
disposition of Majorana fermions in quantum wire networks one can construct a
network with required property. Being far from detailed treatment of Majorana
wire networks in terms of BdG equation, we will focus on finding specific
solutions of the BdG equation metric star graph for zero energy. Here we impose
the vertex boundary conditions providing continuity and current conservation.
General solution of Eq.\re{bdg3} (for Nambu spinor) can be written as
\begin{align}
\Psi^{(j)}(x)=&\mu_{\alpha}^{(j)}\left(\begin{array}{cccc}q^*\\0\\0\\-q\end{array}\right)e^{-\Delta_0
x}+\mu_{\beta}^{(j)}\left(\begin{array}{cccc}0\\q\\q^*\\0\end{array}\right)e^{-\Delta_0
x}\nonumber\\
+&\hat{\mu}_{\alpha}^{(j)}\left(\begin{array}{cccc}q\\0\\0\\-q^*\end{array}\right)e^{\Delta_0
x}+\hat{\mu}_{\beta}^{(j)}\left(\begin{array}{cccc}0\\q^*\\q\\0\end{array}\right)e^{\Delta_0
x}.\label{solE=0}
\end{align}
where $q=1+i$.

Furthermore, we choose the following vertex boundary conditions:
\begin{align}
&\Psi_1^{(1)}(0)=\Psi_1^{(2)}(0)=\Psi_1^{(3)}(0),\label{example6}\\
&\Psi_2^{(1)}(0)+\Psi_2^{(2)}(0)+\Psi_2^{(3)}(0)=0,\label{example7}\\
&\Psi_3^{(1)}(0)+\Psi_3^{(2)}(0)+\Psi_3^{(3)}(0)=0,\label{example8}\\
&\Psi_4^{(1)}(0)=\Psi_4^{(2)}(0)=\Psi_4^{(3)}(0),\label{example9}\\
&\Psi_{1}^{(j)}(L_j)=\Psi_4^{(j)}(L_j),\,\Psi_2^{(j)}(L_j)=\Psi_3^{(j)}(L_j),\nonumber\\
&j=1,2,3,\label{example10}
\end{align}
which provide self-adjointness of the BdG operator on metric star
graph. Eqs.\re{example6} and \re{example9} provides continuity of
wave function, while  Eqs.\re{example7} and \re{example6} lead to
Kirchoff rule at the vertex.

Explicit solutions fulfilling these boundary conditions (for
$L_1=L_2=L_3=L$) can be written as
\begin{align}
&\Psi^{(1,2)}(x)=\left(\begin{array}{c}q^*\\q\\q^*\\-q\end{array}\right)e^{\Delta_0(L-x)}+\left(\begin{array}{c}-q\\q^*\\q\\q^*\end{array}\right)e^{-\Delta_0(L-x)},\nonumber\\
&\Psi^{(3)}(x)=\left(\begin{array}{c}q^*\\-2q\\-2q^*\\-q\end{array}\right)e^{\Delta_0(L-x)}+\left(\begin{array}{c}-q\\-2q^*\\-2q\\q^*\end{array}\right)e^{-\Delta_0(L-x)}.\label{eigfunc1}
\end{align}

Using the same prescription as in the previous section, one can derive vertex
transmission matrix for zero-energy solution which can be written as
\begin{equation}
\bf{T}=-\left({\bf AA'}+{\bf BB'}\right)^{(-1)}\left({\bf
AA''}+{\bf BB''}\right),\label{eq13}
\end{equation}
where
$${\bf
A}'=\left(\begin{array}{cccc}q^*I_N&0&0&0\\0&q^*I_N&0&0\\0&0&qI_N&0\\0&0&0&q^*I_N\end{array}\right),
$$
$${\bf
A}''=\left(\begin{array}{cccc}qI_N&0&0&0\\0&qI_N&0&0\\0&0&q^*I_N&0\\0&0&0&qI_N\end{array}\right),
$$
$${\bf
B}'=\left(\begin{array}{cccc}0&qI_N&0&0\\qI_N&0&0&0\\0&0&0&-q^*I_N\\0&0&-q^*I_N&0\end{array}\right),
$$
$${\bf
B}''=\left(\begin{array}{cccc}0&q^*I_N&0&0\\q^*I_N&0&0&0\\0&0&0&-qI_N\\0&0&-qI_N&0\end{array}\right)
$$
and $\det\left({\bf AA'}+{\bf BB'}\right)\not=0$. Then the vertex transition
matrix can be written as \cite{Bolte}
\begin{equation}
\bf{T}=-\left({\bf AA'}+{\bf BB'}\right)^{(-1)}\left({\bf
AA''}+{\bf BB''}\right).\label{eq13}
\end{equation}
The above approach can be an effective  model for branched Majoana wires with
different topologies and vertex structures. Considering the appropriate vertex
boundary conditions for Eq.\re{bdg3}, one can treat different model
realizations of Majorana wire networks.

\section{Conclusions}\lab{conc}
We have studied the first order Bolgoliubov de Gennes equation on
a metric star graph. The vertex boundary conditions providing the
self-adjointness realization of the BdG operator on a metric star
graph are derived for non-zero and zero energy cases. The
solutions of some special types of the vertex boundary conditions
providing continuity and current conservation are obtained. The
secular equation for finding of eigenvalues is derived. The above
results can be used for modeling of one dimensional branched
superconductors and Majorana wire networks by choosing appropriate
boundary conditions at the branching points.

\end{document}